\documentclass[apj,twocolumn,twocolappendix,numberedappendix]{openjournal}
\usepackage{newtxtext,newtxmath,enumitem,multirow,ctable}
\usepackage[T1]{fontenc}
\usepackage[breaklinks,colorlinks,citecolor=blue,urlcolor=blue]{hyperref}

\newcommand{\gizmourl}{\href{http://www.tapir.caltech.edu/~phopkins/Site/GIZMO.html}{\url{http://www.tapir.caltech.edu/~phopkins/Site/GIZMO.html}}}
\newcommand{\FIREurl}{\href{http://fire.northwestern.edu}{\url{http://fire.northwestern.edu}}}

\newcommand{\orcidauthor}[3]{\author{\href{http://orcid.org/#1}{#2$^{#3}$}}}

\shorttitle{Matters of Momentum}
\shortauthors{Hopkins et al.}

\begin{document}

\title{\vspace{-0.8cm}The Importance of Subtleties in the Scaling of the ``Terminal Momentum'' For Galaxy Formation Simulations\vspace{-1.5cm}}

\orcidauthor{0000-0003-3729-1684}{Philip F. Hopkins}{1,*}
\affiliation{$^{1}$TAPIR, Mailcode 350-17, California Institute of Technology, Pasadena, CA 91125, USA}

\thanks{$^*$E-mail: \href{mailto:phopkins@caltech.edu}{phopkins@caltech.edu}},

\begin{abstract}
In galaxy formation simulations, it is increasingly common to represent supernovae (SNe) at finite resolution (when the Sedov-Taylor phase is unresolved) via hybrid energy-momentum coupling with some ``terminal momentum'' $p_{\rm term}$ (depending weakly on ambient density and metallicity) that represents unresolved work from an energy-conserving phase. Numerical implementations can ensure momentum and energy conservation of such methods, but these require that couplings depend on the surrounding gas velocity field (radial velocity $\langle v_{r} \rangle$). This raises the question of whether $p_{\rm term}$ should also be velocity-dependent, which we explore analytically and in simulations. We show that for simple spherical models, the dependence of $p_{\rm term}$ on $\langle v_{r} \rangle$ introduces negligible corrections beyond those already imposed by energy conservation if $\langle v_{r} \rangle \ge 0$. However, for SNe in some net converging flow ($\langle v_{r} \rangle<0$), naively coupling the total momentum when a blastwave reaches the standard cooling/snowplow phase (or some effective cooling time/velocity/temperature criterion) leads to enormous $p_{\rm term}$ and potentially pathological behaviors. We propose an alternative ``$\Delta$-Momentum'' formulation which represents the differential SNe effect and show this leads to {\em opposite} behavior of $p_{\rm term}$ in this limit. We also consider a more conservative velocity-independent formulation. Testing in numerical simulations, these directly translate to large effects on predicted star formation histories and stellar masses of massive galaxies, explaining  differences between some models and motivating further study in idealized simulations.
\end{abstract}


\maketitle

\section{Introduction}
\label{sec:intro}

Stellar feedback is critical to understanding galaxy formation and regulates the masses and star formation rates (SFRs) and outflows of most galaxies, especially those of Milky Way mass or smaller \citep[][and references therein]{naab.ostriker:2017.galaxy.formation.theory.review,vogelsberger:2020.review.galaxy.form.sims,sales:2022.dwarf.galaxy.simulation.review}. Of different mechanisms, mechanical feedback from SNe is one of, if not the, most important \citep[though see][]{hopkins:rad.pressure.sf.fb,hopkins:fb.ism.prop,tasker:2011.photoion.heating.gmc.evol,kannan:2013.early.fb.gives.good.highz.mgal.mhalo,agertz:2013.new.stellar.fb.model,hopkins:cr.mhd.fire2,ji:fire.cr.cgm,su:2021.agn.jet.params.vs.quenching}. But it has been known for decades that, at achievable resolution for cosmological galaxy-formation simulations of all but the smallest dwarf galaxies, the cooling radii of individual SNe remnants is poorly resolved, so simply coupling the pure ejecta thermal energy and momentum leads to immediate ``overcooling'' and little effect of SNe on galaxy formation (see references above and e.g.\ \citealt{thackercouchman00,marri:2003.mod.sph.cosmo.sims,governato04:resolution.fx}). 

\begin{figure*}
	\centering\includegraphics[width=0.45\textwidth]{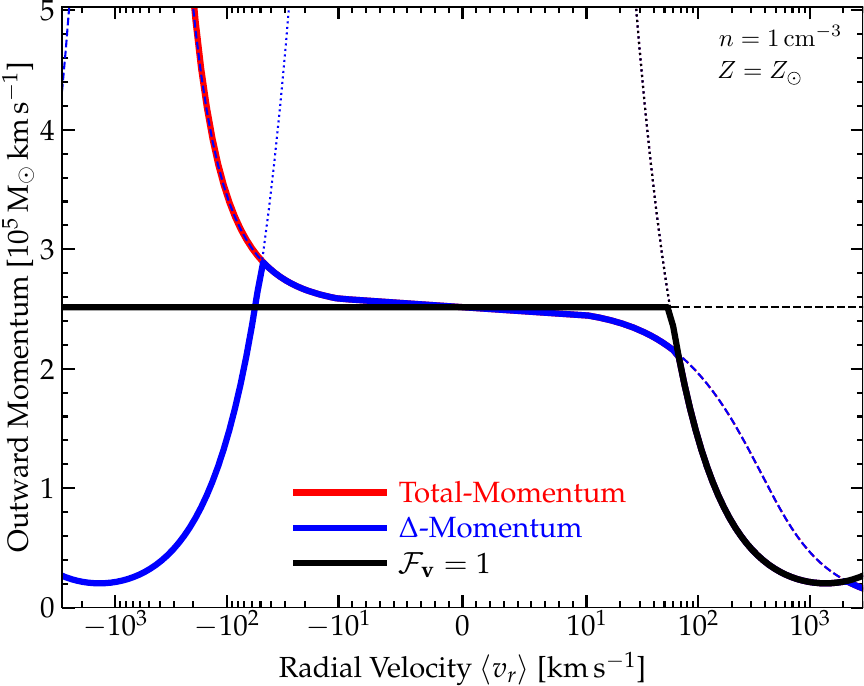} 
	\centering\includegraphics[width=0.47\textwidth]{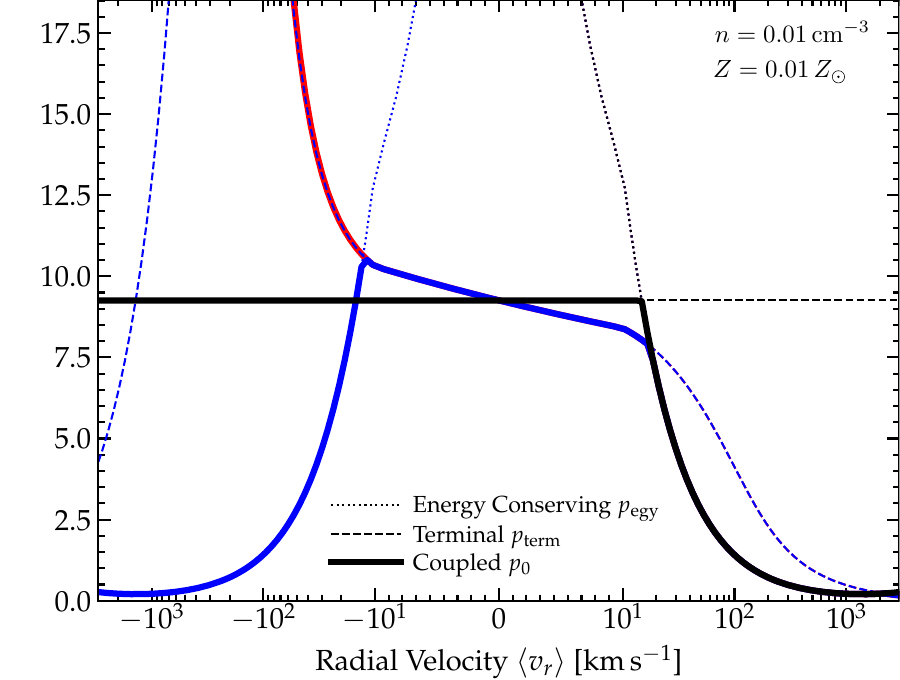} 
	\caption{Momentum associated with different definitions of $\mathcal{F}_{\bf v}$, as a function of the mean radial velocity of gas around a SNe $\langle v_{r} \rangle$. We assume $E_{0}=10^{51}\,{\rm erg}$; $\langle M_{\rm coupled} \rangle \sim 10^{6}\,{\rm M_{\odot}}$ (the results are insensitive so long as this is larger than the ``cooling mass'' $\sim 2000\,{\rm M_{\odot}}$); and $n=1\,{\rm cm^{-3}}$, $Z=Z_{\odot}$ ({\em left}) or $n=0.01\,{\rm cm^{-3}}$, $Z=0.01\,Z_{\odot}$ ({\em right}), with $\beta_{\Lambda}\approx0.5$ for Bremsstrahlung cooling. We compare three choices of $\mathcal{F}_{\bf v}$: (1) that given by the ``total momentum change'' between pre-SNe and asymptotic post-SNe states (\S~\ref{sec:total.mom}), (2) the alternative ``$\Delta$-Momentum'' (difference in non-linear states with and without a SNe; \S~\ref{sec:delta.mom}), or (3) simply taking $\mathcal{F}_{\bf v}\rightarrow 1$. For each we show the energy-conserving $p_{\rm egy}$ or $\Delta p_{\rm egy}$, ``terminal'' momentum $p_{\rm term}$, and minimum (coupled $p_{0}$; Eq.~\ref{eqn:p.coupled.summary}). For $\langle v_{r} \rangle \gtrsim -10\,{\rm km\,s^{-1}}$, the behaviors are similar, but they differ dramatically in converging flows.
	\label{fig:pterm.compare}}
\end{figure*}

In the last decade, an increasingly popular method (at least in ``semi-resolved-ISM'' simulations with cold neutral media and mass resolution $\lesssim 10^{6}\,{\rm M_{\odot}}$) to address this dilemma is to implement mechanical feedback models which couple both thermal and kinetic energy/momentum \citep[see e.g.][]{hopkins:fb.ism.prop,hopkins:2013.fire,hopkins:fire2.methods,hopkins:fire3.methods,kimm.cen:escape.fraction,agertz:sf.feedback.multiple.mechanisms,rosdahl:2016.sne.method.isolated.gal.sims,hu:2017.rad.fb.model.photoelectric,read:2017.shallow.mgal.mhalo,lupi:2017.gizmo.galaxy.form.methods,kim:tigress.ism.model.sims,marinacci:2019.smuggle.models,lahen:2020.griffin.sims,agertz.2021:vintergatan.sims,martin.alvarez:radiation.crs.galsim.similar.conclusions.fire}. Most of these algorithms reflect some variant of those in \citet{hopkins:fb.ism.prop,hopkins:2013.fire}, where the momentum coupled is scaled with the resolution/coupling scale in order to represent the ``PdV'' work done by the hot SNe bubble in whatever energy-conserving phase is un-resolved -- the so-called ``terminal momentum'' $p_{\rm term}$. This terminal momentum has since been extensively studied in orders-of-magnitude higher resolution simulations including the effects of different ambient density, metallicity, magnetic fields, SNe energy, multiple clustered SNe, pre-SNe feedback, and surrounding gas turbulence \citep{hennebelle.iffrig.multi.sne.sims,iffrig:sne.momentum.magnetic.no.effects,martizzi:sne.momentum.sims,walch.naab:sne.momentum,kim.ostriker:sne.momentum.injection.sims,haid:snr.in.clumpy.ism,gentry:clustered.sne.momentum.enhancement,gentry:sne.momentum.boost,ohlin:2019.sne.idealized.in.turb.terminal.momentum,elbadry:2019.superbubble.conduction.solution.modifications,zhang:2019.sne.in.turb.clouds,pittard:2019.clustered.sne.less.efficient.momentum.injection.eventually,steinwandel:2019.magnetic.bouyancy.galactic.scales}, which in general have found it varies only weakly with these parameters, and agrees surprisingly well with classic, simple analytic models  \citep[e.g.][]{chevalier:1974.spherical.sne.blastwave}. In the meantime, numerical improvements to these algorithms have been developed \citep{hopkins:sne.methods,hopkins:fire3.methods} which allow for such coupling to manifestly ensure mass, momentum, and energy conservation as well as the translational, Galilean, and rotational symmetries intrinsic to the problem and avoid imprinting spurious effects such as preferred grid directions.

As discussed in \citet{hopkins:fire3.methods}, ensuring exact total energy conservation in the SNe coupling operation {\em requires} that the effective total momentum coupled in the low-resolution limit depend, at least in some cases, on the surrounding velocity field -- i.e.\ $p_{\rm term}$ cannot remain very weakly velocity-dependent for arbitrary surrounding velocity fields. This becomes important in a regime which most of the idealized simulations above did not explore, namely when there is a large {\em net} velocity divergence (inflow/outflow or converging/diverging flow $\langle v_{r} \rangle$) around a given SNe (although this effect was noted for SNe in pre-existing outflows in some studies above, such as \citealt{pittard:2019.clustered.sne.less.efficient.momentum.injection.eventually}). Indeed most of the high-resolution studies cited which did not assume an initially stationary ambient medium considered a uniformly turbulent medium, so the net inflow+outflow around a median SNe cancel and the effect discussed here is small \citep[see discussion in][]{hopkins:sne.methods}.  As a result, the vast majority of the prescriptions cited above and used in galaxy-formation simulations above simply adopt a constant $p_{\rm term}$, regardless of the true solution velocity dependence or energy conservation considerations. 

In this paper, we therefore consider this question in more detail, and show that an important ambiguity arises when SNe explode in strongly-converging flows. While this may not be relevant for the ``median'' SNe, we show that it can have a large effect on galaxy formation (at typically-adopted numerical resolution) if standard analytic models for the terminal momentum and blastwave evolution are extended to media with non-vanishing $\langle v_{r} \rangle$. We show that this can explain some surprisingly large differences between galaxy formation simulations with seemingly similar physics, and motivate further idealized (high-resolution) simulations for the regime of strong converging flows. In the process, we present a more simplified and flexible version of the manifestly-conservative SNe algorithm from \citet{hopkins:sne.methods,hopkins:fire3.methods}, which we make public in the code GIZMO.

\section{Supernova Coupling}
\label{sec:sne}

\subsection{Basic Equations}

Consider a simulation where a mechanical feedback ``event'' occurs and we wish to couple a combination of energy and momentum to the surrounding cells following \citet{hopkins:sne.methods} and \citet{hopkins:fire3.methods} as detailed in Appendix~\ref{sec:methods}. In the frame of the source (star) $a$, surrounding gas cells $b$ are assigned some scalar products (ejecta mass, metals, thermal or cosmic ray or radiation energy) plus some outward momentum 
\begin{align}
\Delta {\bf p}_{ba} \equiv \bar{\bf w}_{ba} p_{0}
\end{align}
where $\bar{\bf w}_{ba}$ is some dimensionless vector weight function (carefully chosen to manifestly ensure isotropy and translation, Galilean, and rotation invariance plus mass, momentum, and energy conservation; see \S~\ref{sec:methods:eqns}), and $p_{0} = \sum_{b} |\Delta {\bf p}_{ba}|$ is the total injected/coupled radial/outward momentum (in the frame of the star). 
Following \S~\ref{sec:methods:egycon}, this can be written as: 
\begin{align}
\label{eqn:p.coupled.summary} p_{0} &\equiv \psi\, \chi\, (2\,\epsilon\, m_{\rm ej})^{1/2} = {\rm MIN}[ p_{\rm egy}\ , \ p_{\rm term}]
\end{align}
where $\epsilon$ (defined precisely in Eq.~\ref{eqn:epsilon}) is roughly the kinetic energy of the ejecta (plus star-gas relative motions), 
$p_{\rm term}$ is some limiting or ``terminal'' momentum (so $\chi \equiv {\rm MIN}[1,\,p_{\rm term}/p_{\rm egy}]$), and 
\begin{align}
p_{\rm egy} \equiv \psi\,p_{\rm ej} = \psi\, (2\, \epsilon\, m_{\rm ej})^{1/2}  \approx \psi\, p_{\rm ej}
\end{align}
is the momentum one would obtain in the strictly energy-conserving limit (in terms of the effective ejecta momentum $p_{\rm ej} = m_{\rm ej}\,v_{\rm ej} \equiv (2\,\epsilon\, m_{\rm ej})^{1/2}$). As shown in \S~\ref{sec:methods:egycon}, following a strict energy-conserving Sedov-type solution -- i.e.\ setting the final kinetic energy of all cells equal to their initial kinetic energy plus the injected kinetic energy with no radiation/cooling losses -- determines the dimensionless function $\psi \equiv \beta_{2}^{-1}\,[\sqrt{\beta_{2}+\beta_{1}^{2}} - \beta_{1}]$ in terms of\footnote{Here ${\bf w}_{ba}^{\prime} \equiv \bar{\bf w}_{ba}/ (1+|\bar{\bf w}_{ba}| m_{\rm ej}/m_{b})$ (see \S~\ref{sec:methods:egycon}).} $\beta_{1} \equiv v_{\rm ej}^{-1} \sum_{b} {\bf v}_{ba} \cdot {\bf w}^{\prime}_{ba}  \equiv \langle v_{r} \rangle_{a} / v_{\rm ej}$ (roughly, the mean surrounding inflow/outflow velocity $\langle v_{r} \rangle$ relative to the ejecta velocity) and $\beta_{2}\equiv m_{\rm ej} \sum_{b} | {\bf w}^{\prime}| \,m_{b}^{-1} \equiv m_{\rm ej} / \langle M_{\rm coupled} \rangle $ (roughly, the ratio of ejecta mass to the effective ambient gas mass over which the ejecta is coupled, see \S~\ref{sec:methods:mcoupled}).

\subsection{Limiting Behaviors}
\label{sec:methods:limits}

Exact solutions for $p_{\rm egy}$ and $p_{0}$ as a function of $\langle v_{r}\rangle$ and $\langle M_{\rm coupled} \rangle$ (equivalently, $\beta_{1}$ and $\beta_{2}$) are shown in Fig.~\ref{fig:pterm.compare}. But in summary, the coupled momentum above (Eq.~\ref{eqn:p.coupled.summary}, or derived rigorously in Eq.~\ref{eqn:p0}) has several limiting behaviors. 
\begin{itemize}
\item If $\chi < 1$ (i.e.\ $p_{\rm term} < p_{\rm egy} \equiv \psi\,(2\,\epsilon\,m_{{\rm ej}})^{1/2}$), then 
\begin{align}
p_{0} \rightarrow p_{\rm term}
\end{align}
by construction. This is the ``terminal momentum'' limit, where the momentum coupled is ``capped'' by $p_{\rm term}$ and so will depend on how we define the $p_{\rm term}$ below.

\item For $\chi=1$ ($p_{\rm egy} > p_{\rm term}$), if $\beta_{1}^{2}/\beta_{2} \ll 1$, then $\psi \rightarrow \beta_{2}^{-1/2} = \sqrt{\langle M_{\rm coupled} \rangle/ m_{\rm ej}}$, so the total coupled momentum is 
\begin{align}
p_{0} \rightarrow \psi \, p_{\rm ej} \rightarrow \sqrt{2\,\epsilon\, \langle M_{\rm coupled} \rangle}\ .
\end{align}
This is the expectation for an energy-conserving solution in a homogeneous, non-moving medium -- in other words, the relative velocities (appearing in $\beta_{1} = \langle v_{r} \rangle/ v_{\rm ej}$) are sufficiently small to be negligible. This can become quite large if $\langle M_{\rm coupled} \rangle$ is large -- i.e.\ at low resolution. In that case, one will generally hit the terminal momentum limit instead: in other words, the energy-conserving phase will be unresolved.

\item For $\chi=1$ ($p_{\rm egy} > p_{\rm term}$), if $\beta_{1}^{2}/\beta_{2} \gg 1$, then the relative star-gas inflow/outflow velocity $\langle v_{r} \rangle$ is not negligible. The behavior then depends on the sign of $\beta_{1} \propto \langle v_{r} \rangle$. If $\langle v_{r} \rangle > 0$ (the gas is net receding from the star), $\psi \rightarrow 1/2\beta_{1}$ and 
\begin{align}
p_{0} &\rightarrow \frac{v_{\rm ej}}{2\langle v_{r} \rangle}\,m_{\rm ej}\,v_{\rm ej} = \frac{\epsilon}{\langle v_{r} \rangle} \\
\nonumber &=\sqrt{2\,\epsilon\, \langle M_{\rm coupled} \rangle} \sqrt{ \frac{\epsilon }{  2 \langle M_{\rm coupled} \rangle \, \langle v_{r} \rangle^{2}}}\ .
\end{align}
Note that whether we are in this regime depends on the ratio $\beta_{1}^{2}/\beta_{2} = \langle M_{\rm coupled} \rangle\,\langle v_{r} \rangle^{2} / m_{\rm ej}\,v_{\rm ej}^{2} \sim {\rm KE}^{\rm coupled}_{r} / {\rm KE}_{\rm ej}$, i.e.\ just the ratio of the pre-existing radial kinetic energy of the background gas to the ejecta kinetic energy. When this is large, the coupled momentum is suppressed (relative to the energy-conserving solution in a non-moving background) by the root of this factor (i.e.\ $|\psi| \ll 1$), to enforce total energy conservation, since the coupled kinetic energy is a non-linear function of the coupled momentum. 
We see in Fig.~\ref{fig:pterm.compare} that this generally limits the coupled momentum to be less than the naive $p_{\rm term}$ one would obtain in a non-moving background when $\langle v_{r}\rangle \gg 10-100\,{\rm km\,s^{-1}}$. Coupling the ``full'' $p_{\rm term}$ would then require $>10^{51}\,{\rm erg}$ of increased kinetic energy.

\item If $\chi=1$ ($p_{\rm egy} > p_{\rm term}$) and $\beta_{1}^{2}/\beta_{2} \gg 1$, and $\langle v_{r} \rangle < 0$ (the gas is net infalling towards the star), then $\psi \rightarrow -2\,\beta_{1}/\beta_{2}$ and 
\begin{align}
p_{0} \rightarrow -2\,\langle M_{\rm coupled} \rangle \, \langle v_{r} \rangle
\end{align} 
which by definition means $|p_{0} | \gg p_{\rm ej}$. 
In this limit, from an energetic point of view, there is a large energy reservoir in the inflow kinetic energy ($\langle M_{\rm coupled} \rangle \langle v_{r} \rangle^{2} \gg \epsilon$). There is, therefore, always sufficient energy in principle to ``turn around'' and flip the initial inflow $\langle v_{r} \rangle_{\rm initial} < 0$ to outflow with similar speed $\langle v_{r} \rangle_{\rm final} \sim |\langle v_{r}\rangle_{\rm initial}|$. 
In practice (Fig.~\ref{fig:pterm.compare}), since this can imply a very large $p_{\rm egy}$ ($|\psi| \gg 1$), then if $p_{\rm term}$ is {\em also} assumed to become large in this limit, the coupled momentum can become extremely large. If $p_{\rm term}$ remains relatively small, however, then this limit cannot be reached, and one will hit the terminal momentum limit instead.
\end{itemize}

\subsection{General Scalings for the Terminal Momentum}
\label{sec:pterm.general.subfunctions}


Given these behaviors and limits, for SNe,\footnote{Since we are interested in SNe (which carry most of the energy and momentum), we follow \citet{hopkins:fire3.methods} and for non-SNe (e.g.\ stellar mass-loss) simply set $p_{\rm term} \rightarrow p_{\rm ej}$.} the important ``sub-grid physics'' is contained, by construction, in the function $p_{\rm term}$, which is some function of ambient gas properties, themselves estimated self-consistently from the surrounding cells (defined precisely in \S~\ref{sec:methods:eval}). For simplicity, assume a separable function of the form:
\begin{align}
p_{\rm term} &= p_{t,\,0}\,\mathcal{F}_{\mathcal{E}}(\mathcal{E})\mathcal{F}_{n}(\langle n\rangle)\,\mathcal{F}_{Z}(\langle Z \rangle)\,\mathcal{F}_{\bf v}(\langle {\bf v}\rangle) 
\end{align}
in terms of the total SNe-frame ejecta energy $\mathcal{E}$, and gas density $\langle n\rangle$, metallicity $\langle Z\rangle$, and velocity field $\langle {\bf v} \rangle = {\bf v}_{\rm gas}({\bf x}-{\bf x}_{\ast}) - {\bf v}_{\ast}$. 
Our ``reference'' scalings assumed in \citet{hopkins:fire3.methods} are: $\mathcal{F}_{\mathcal{E}} = \mathcal{E}/10^{51}\,{\rm erg}$; $\mathcal{F}_{n}(\tilde{n} \equiv \langle n \rangle/{\rm cm^{-3}}) = 2.63$ for $\tilde{n}<0.001$ and $=\tilde{n}^{-0.143}$ for $\tilde{n}\ge 0.001$; $\mathcal{F}_{Z}(\tilde{z} \equiv Z/Z_{\odot}) = 2$ for $\tilde{z}<0.01$, $=\tilde{z}^{-0.18}$ for $0.01 \le \tilde{z} \le 1$ and $=\tilde{z}^{-0.12}$ for $\tilde{z}>1$; and $p_{t,\,0}/{\rm M_{\odot}\,km\,s^{-1}} = (f_{\rm kin}^{0})^{1/2}\,4.8\times10^{5} \approx 2.5\times10^{5}$. We can also write $p_{\rm term} = (\mathcal{E}/v_{\rm term})\,\mathcal{F}_{{\bf v}}$ with $v_{\rm term} \approx 200\,{\rm km\,s^{-1}}\,\mathcal{F}_{n}^{-1}\,\mathcal{F}_{Z}^{-1}$. This $v_{\rm term}$ is the ``terminal velocity,'' usually described as the shock velocity at which a remnant should begin to rapidly cool and transition from a Sedov-Taylor to snowplow-type phase.

The scalings for $p_{t,\,0}$, $\mathcal{F}_{\mathcal{E}}$, $\mathcal{F}_{n}$, and $\mathcal{F}_{Z}$ are relatively well-understood: they can be derived both analytically and empirically from numerical simulations as the result of where SNe explosions will efficiently cool (see references in \S~\ref{sec:intro}), and are weak. The poorly-understood scaling, and our main focus, is the function $\mathcal{F}_{\bf v}$.

\begin{figure}
	\centering\includegraphics[width=0.95\columnwidth]{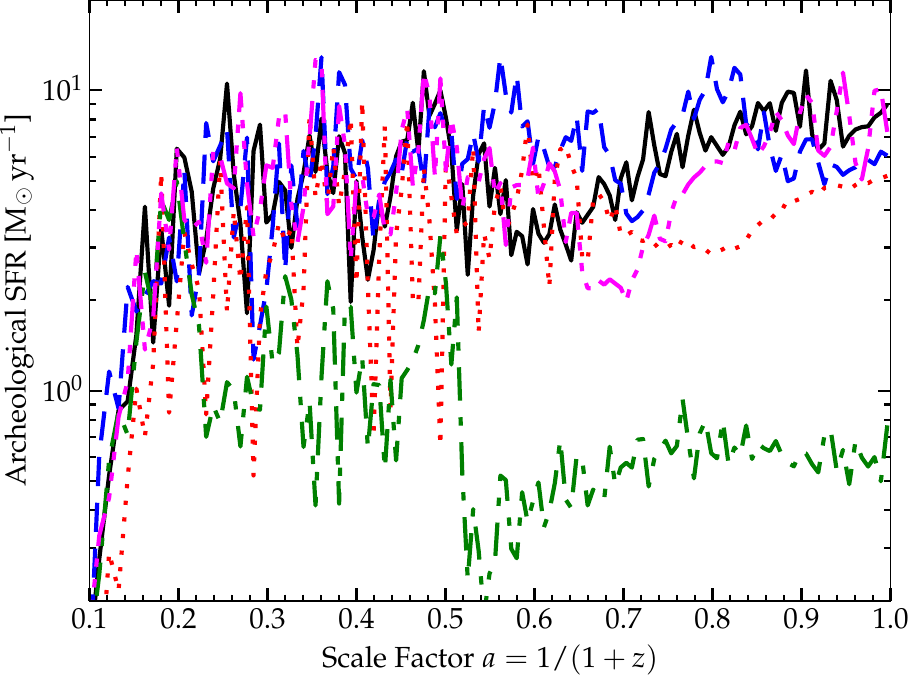} 
	\centering\includegraphics[width=0.95\columnwidth]{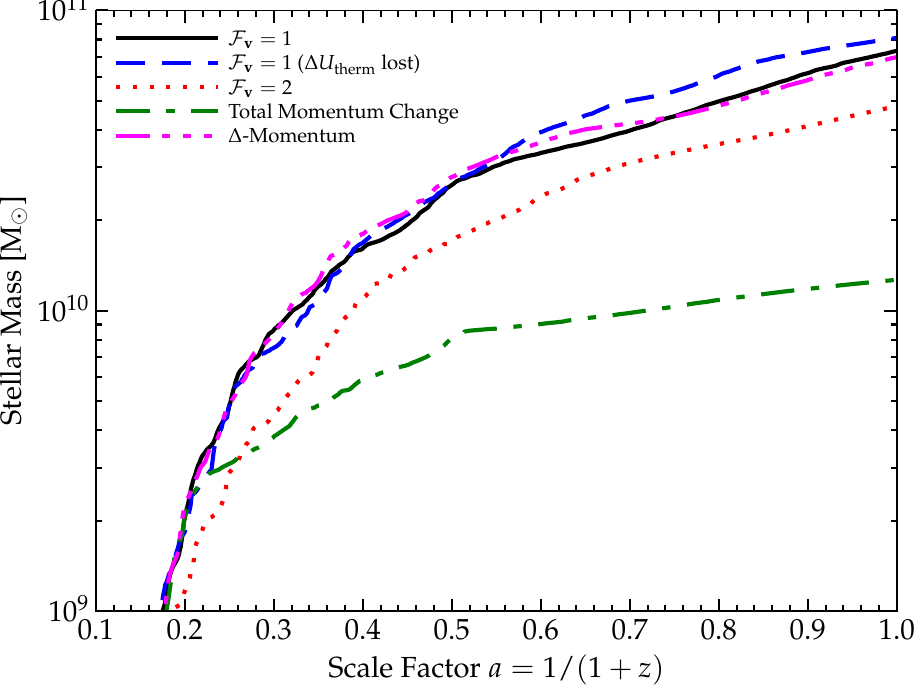} 
	\caption{Effects of different terminal momentum choices in cosmological simulations (\S~\ref{sec:effects}). We compare otherwise identical cosmological simulations of a Milky Way-mass halo at $z=0$ with FIRE-3, varying $\mathcal{F}_{\bf v}$, at intermediate resolution ($\langle \Delta m \rangle \sim 6\times10^{4}\,{\rm M_{\odot}}$) We show the archeological SF history and stellar mass growth. At this resolution individual SNe cooling radii are unresolved, so systematically changing the terminal momentum via $\mathcal{F}_{\bf v}$ directly renormalizes the SFR. Using the ``total momentum change'' (\S~\ref{sec:total.mom}) formulation leads to a very strong suppression of SF at late times (once the potential becomes deep enough that with fixed $\mathcal{F}_{\bf v} \sim 1$, inflows would become stronger). Using a fixed $\mathcal{F}_{\bf v}=1$ or $\mathcal{F}_{\bf v}$ from the ``$\Delta$-Momentum'' formulation (\S~\ref{sec:delta.mom}) give similar results.
	\label{fig:m12i.compare}}
\end{figure}

\section{How Should the Terminal Momentum Depend on Velocity?}
\label{sec:Fv}

\subsection{A First Guess: The ``Total'' Change in Momentum}
\label{sec:total.mom}

Of course, on the size scale of the SNe blastwave and coupling to the gas, the ``true'' velocity field could be infinitely complex, and ultimately understanding $\mathcal{F}_{\bf v}$ requires numerical simulations. But we wish to gain some intuition with simple models. If we write the smooth (on this scale) large-scale velocity field as ${\bf v} \approx \langle {\bf v} \rangle + \langle \nabla {\bf v} \rangle \cdot {\bf x} + \mathcal{O}({\bf x}^{2})$ (in the frame of the star), and decompose $\nabla {\bf v}$ into the usual trace-free/deviatoric shear, rotation, and expansion/compression ($\nabla \cdot {\bf v} \sim \langle v_{r} \rangle / r$) terms, we can gain some insight. Since our SNe solutions assume spherical symmetry, the only of these terms which should enter is the $\nabla \cdot {\bf v} \rightarrow r^{-2}\,\partial_{r}(r^{2}\,v_{r})$, i.e.\ $v_{r}$ term.\footnote{In more detail, in \citet{hopkins:sne.methods} and the Appendices of \citet{hopkins:fire3.methods}, we show that the mean term $\langle {\bf v} \rangle$ should not change the spherically-averaged growth of a blastwave (beyond accounting for the term in the effective energy $\mathcal{E}$ which appears in Appendix~\ref{sec:methods}). Moreover it should be obvious that the pure rotation/circulation part of $\nabla {\bf v}$ (velocities transverse to the ejecta) should not strongly influence the ejecta evolution. The deviatoric shear could be important in reality (one could imagine e.g.\ a blastwave compressed in some ``midplane'' direction along which $v_{r} < 0$ but venting out of the polar direction where $v_{r} > 0$), but this would require modeling effects beyond spherically symmetric injection. So to begin we will consider just the radial velocity (inflow/outflow) around the SNe $v_{r}$.} In \S~\ref{sec:analytic.fv} we re-derive the usual analytic terminal momentum argument by considering the radius at which a spherically-symmetric blastwave would have its cooling time fall below its dynamical time, now allowing for a uniform radial velocity $\langle v_{r} \rangle$ in the frame of the star, and show the result can be approximated as $\mathcal{F}_{\bf v} \approx 1/(1 + \langle v_{r} \rangle/v_{\rm term})$. 

For $\langle v_{r} \rangle \ge 0$, this gives completely reasonable behavior (Fig.~\ref{fig:pterm.compare}). At small $\langle v_{r} \rangle \rightarrow 0$, $\mathcal{F}_{\bf v}\rightarrow 1$, as it should, and for $\langle v_{r} \rangle \gg v_{\rm term}$, $\mathcal{F}_{\bf v} \rightarrow v_{\rm term}/\langle v_{r}\rangle$ and $p_{\rm term} \rightarrow E_{0}/\langle v_{r} \rangle$. But this is the same behavior as $p_{\rm egy}$ at large $\langle v_{r}\rangle > 0$ (because both reflect the same total energy; \S~\ref{sec:methods:limits}). So it is functionally identical to assume $\mathcal{F}_{{\bf v}} = 1$ for all $\langle v_{r} \rangle \ge 0$. 

The important difference arises for negative $\langle v_{r} \rangle < 0$, i.e.\ net infall towards the SNe. The simple expression above gives $\mathcal{F}_{\bf v} \rightarrow \infty$ for $\langle v_{r} \rangle \le v_{\rm term}$, and while the more detailed expressions derived in \S~\ref{sec:analytic.fv} do not formally diverge, they give effectively the same result, with $\mathcal{F}_{\bf v} \sim |\langle v_{r} \rangle / v_{\rm term}|^{14}$ or so for $\langle v_{r} \rangle \ll -v_{\rm term}$. As $p_{\rm term} \rightarrow \infty$, then one never reaches the ``terminal'' momentum, because the inflowing gas velocity ensures the post-shock temperature is so hot that cooling is remains inefficient. 

As shown in Fig.~\ref{fig:pterm.compare}, this means that for large negative/inflow $\langle v_{r} \rangle$, the coupled $p_{0} = {\rm MIN}[ p_{\rm egy},\, p_{\rm term}]$ will be given by the strictly energy-conserving $p_{\rm egy}$. But recall from \S~\ref{sec:methods:limits} that in this limit at low resolution, $p_{\rm egy} \rightarrow 2\,\langle M_{\rm coupled}\rangle |\langle v_{r} \rangle|$ also becomes very large -- so large, in fact, that the inflow would always be halted and reversed. I.e.\ the inflow  energy-loading of the region de-facto becomes an outflow energy-loading, {\em independent} of the actual SNe energy $\epsilon$ injected (even if this becomes vanishingly small).

This is not some unique artifact of assumptions we adopted: we obtain the same qualitative result if we instead approximate the blastwave as occurring in a Hubble flow, or isothermal inflow/wind solution, or even as a one-dimensional surface with some external inflow \citep[see solutions in][]{ostrikermckee:blastwaves}; whether we consider the shell approximation or a homogenous sphere, and whether we consider the energy injection as impulsive or continuous over the timestep; whether we consider the ambient medium to be infinite, or strictly limited to the mass $\langle M_{\rm coupled} \rangle$; or whether we vary the cooling curve ($\beta_{\Lambda}$ in \S~\ref{sec:analytic.fv}). The reason is simple: in this limit of a spherically-symmetric inflow with inflow kinetic energy much larger than ejecta energy ($\langle M_{\rm coupled} \rangle \langle v_{r} \rangle^{2} \gg E_{0}$), the problem transitions fundamentally from a ``blastwave'' to an ``accretion shock.'' Our analytic solutions are qualitatively completely reasonable when we realize they are the standard solutions for an accretion shock, which when $\langle v_{r} \rangle \ll - v_{\rm term}$ corresponds to post-shock cooling times which are long (e.g.\ ``hot'' halos) and so the shocked region grows as the mass supply continues \citep{rees:1977.tcool.tdyn.vs.mhalo}.

\subsection{An Alternative: The Change in Total Momentum Relative to Evolution Without Feedback}
\label{sec:delta.mom}

The ``total momentum'' approach to calculating $p_{\rm term}$ or $\mathcal{F}_{\bf v}$ in \S~\ref{sec:total.mom} therefore leads to a divergence in the coupled momentum $p_{0}$ at large negative/inflow $\langle v_{r} \rangle \ll 0$ and low resolution (Fig.~\ref{fig:pterm.compare}), causing inflow to become outflow independent of the actual SNe energy coupled. But the discussion there (and detailed derivation in \S~\ref{sec:analytic.fv}) also highlights one problematic aspect of that derivation for $\langle v_{r} \rangle \ll 0$: by applying the ``final'' momentum (energy-conserving or terminal) in this limit, we are essentially ``jumping ahead'' of the hydro/MHD solver. If, after all, $\langle v_{r} \rangle \ll -v_{\rm term}$, then there is (by definition) sufficient energy in the {\em resolved} simulation cells to heat themselves above the threshold where cooling becomes inefficient. So the solution should be (in principle) resolved. Thus while not technically wrong or unphysical (or in violation of any conservation law), this ``jumping ahead'' can be problematic for several reasons: (1) we are making many approximations (like spherical symmetry, an idealized cooling function, no other sources/sinks, a single blastwave in isolation, homogeneous and isotropic density/metallicity/velocity fields, no radiation or cosmic rays or other forces, etc.) which the actual surrounding gas may not obey, (2) it is possible that this ``jumping ahead'' could effectively jump to a solution further in time than the actual code timestep would allow (if e.g.\ the gas on either side of the SNe should not be able to ``cross'' and shock yet in the timestep during which the SNe is coupled), and (3) it violates the spirit of a well-posed sub-grid model, in that such should represent the result of only un-resolved processes but ``trust'' the code for dealing with resolved processes. 

A potential alternative is instead to apply the ``$\Delta$-Momentum'': specifically, instead of calculating the final momentum injected in a SNe $p_{0}$ as the difference between the initial (pre-SNe) code momentum state $p_{\rm initial}$ and some analytic ``post-evolution'' blastwave/shock state $p_{\rm final}$ ($p_{0} = p_{\rm final} - p_{\rm initial}$), we can define the $\Delta$-Momentum $\Delta p$ as the difference between these states {\em relative to the solution where there is no SNe at all} (mathematically $\Delta p \equiv (p_{\rm final}-p_{\rm initial})_{E_{0}=E_{0}} -  (p_{\rm final}-p_{\rm initial})_{E_{0}\rightarrow 0}$), in order to identify how said SNe uniquely modifies the state. These solutions are derived in \S~\ref{sec:analytic.fv:delta}. 

Fig.~\ref{fig:pterm.compare} compares the behavior of the naive final momentum coupled $p_{0}$ and the $\Delta$-Momentum $\Delta p$ for realistic simulation conditions. For both the energy-conserving and terminal limit, the $\Delta$-Momentum is identical to the ``normal'' final momentum we would have coupled above, $\Delta p = p_{0}$, for $\langle v_{r} \rangle \ge 0$. But for $\langle v_{r} \rangle < 0$, there is no longer any divergence in momentum coupled, $\Delta p$ vanishes as $E_{0}\rightarrow 0$, and for very large $\langle v_{r} \rangle \ll -v_{\rm term}$, $\Delta p \propto 1/|\langle v_{r} \rangle|$ is a decreasing function of $|\langle v_{r} \rangle|$ (it no longer grows without limit). But we should stress that since the equations being represented here are fundamentally non-linear, it is by no means obvious that the sum of the ad-hoc defined $\Delta$-Momentum plus subsequent resolved solution evolution in-code will actually give the same non-linear answer as a direct solution at much higher resolution.

Given this ambiguity and the range ``bracketed'' by the above final momentum and $\Delta$-momentum, an even simpler alternative is just to always take $\mathcal{F}_{{\bf v}}=1$. We see in Fig.~\ref{fig:pterm.compare} that for $\langle v_{r} \rangle \ge 0$, this is basically identical to either of these more sophisticated approaches, but for $\langle v_{r} \rangle \lesssim -v_{\rm term}$, it is ``in between.''

\begin{figure}
	\centering\includegraphics[width=0.95\columnwidth]{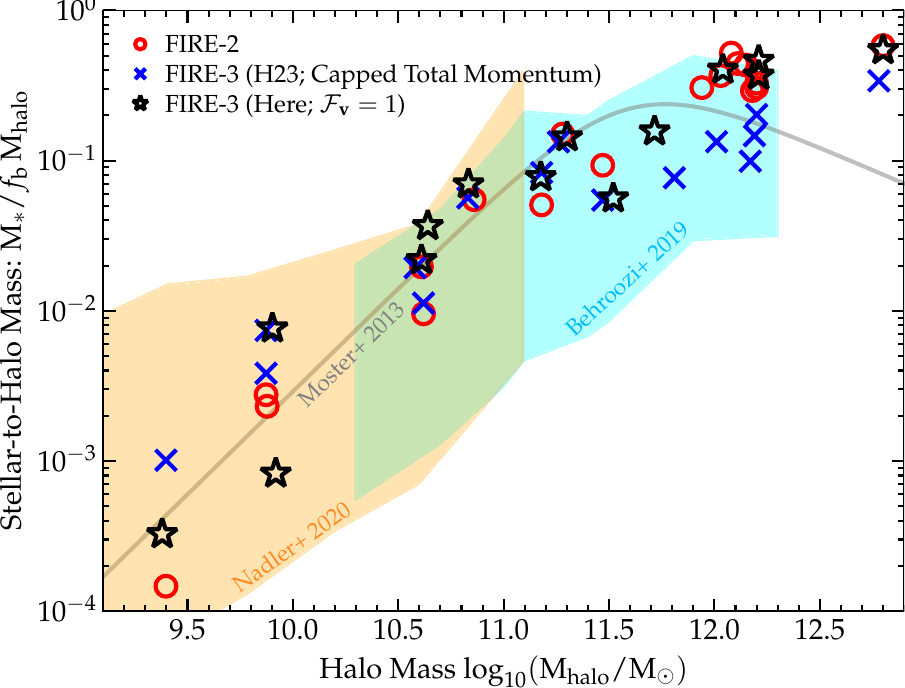} 
	\caption{Comparison of the stellar mass-halo mass relation at $z=0$ from cosmological simulations (\S~\ref{sec:effects}). We plot total stellar mass (within $<20\,$kpc of the halo center) versus halo virial mass for each primary galaxy in a zoom-in region, for identical initial conditions simulated with FIRE-3 adopting the SNe treatment here with $\mathcal{F}_{\bf v}=1$; or using the treatment assumed in \citet[][H23]{hopkins:fire3.methods}, which was similar to the ``Total Momentum Change'' formula for $\mathcal{F}_{\bf v}$ with a ``cap'' at $\mathcal{F}_{\bf v} \lesssim |\bar{\bf w}_{ba}|^{-1/2} \sim $\,a few; and FIRE-2 \citep{hopkins:sne.methods}, which adopted $\mathcal{F}_{\bf v}=1$ but a different and not strictly energy-conserving SNe coupling, different cooling physics, stellar evolution tables, yields, treatment of OB/AGB mass-loss, UV background, and star formation criteria. Details of the differences between SNe treatments are in \S~\ref{sec:compare}. These are hydrodynamics-only (no MHD or cosmic rays), no-AGN, low-resolution simulations (resolution $\sim (300,\,10^{4},\,6\times10^{4},\,3\times10^{5})\,{\rm M_{\odot}}$ for $M_{\rm halo}\sim (10^{10},\,10^{11},\,10^{12},\,10^{13})\,{\rm M_{\odot}}$), but we show the abundance-matching relation from \citet{moster:2013.abundance.matching.sfhs} and the $2\sigma$ scatter from \citet{behroozi:2019.sham.update,nadler:2020.abundance.matching.w.lmc.fx} for reference. The different H23 and FIRE-2 predictions for stellar masses in massive ($M_{\rm halo} \gg 10^{11}\,{\rm M_{\odot}}$) galaxies appears to be dominated by the implicit difference in $\mathcal{F}_{\bf v}$. 
	\label{fig:mstar.mhalo}}
\end{figure}

\section{Effects in Galaxy Formation Simulations}
\label{sec:effects}

We now consider the effects of these choices in cosmological galaxy formation simulations. The specific simulations we consider below were run as part of the Feedback In Realistic Environments (FIRE)\footnote{\FIREurl} project, specifically using the FIRE-3 version of the code; all details of the methods are described in \citet{hopkins:fire3.methods}. Briefly, the simulations use the code {\small GIZMO} \citep{hopkins:gizmo},\footnote{A public version of {\small GIZMO} is available at \gizmourl} with hydrodynamics solved using the mesh-free Lagrangian Godunov MFM method. Both hydrodynamic and gravitational (force-softening) spatial resolution are set in a fully-adaptive Lagrangian manner; mass resolution is fixed. The simulations include cooling and heating from a meta-galactic background and local stellar sources from $T\sim10-10^{10}\,$K; SF in locally self-gravitating and Jeans-unstable gas; and stellar feedback from OB \&\ AGB mass-loss, SNe Ia \&\ II, and multi-wavelength photo-heating and radiation pressure; with inputs taken directly from stellar evolution models. We compare a set of fully-cosmological zoom-in simulations run from initial redshifts $z\sim 100$ down to $z=0$ with fiducial resolution surrounding a ``target'' halo of interest. Here, we keep the physics and numerical methods identical to those in \citet{hopkins:fire3.methods}, and vary only mechanical coupling, using the exact numerical prescription from Appendix~\ref{sec:methods}, with the important physical variations contained in the function $\mathcal{F}_{{\bf v}}$. 

Fig.~\ref{fig:m12i.compare} compares the history of a typical Milky Way-mass galaxy ({\bf m12i} from \citealt{hopkins:2013.fire}), for several choices of $\mathcal{F}_{{\bf v}}$. This simulation has a modest, but common for ``zoom-in'' cosmological simulations, mass resolution of $\langle m_{b} \rangle \sim 6\times10^{4}\,{\rm M_{\odot}}$, which as noted in the Appendices gives $\langle M_{\rm coupled} \rangle \sim (2-4)\times10^{5}\,{\rm M_{\odot}}$, so as a result is almost always in the ``terminal momentum'' limit. We therefore see the expected result from feedback-regulated SF, consistent with many previous more systematic studies with the FIRE simulations \citep[e.g.][]{hopkins:rad.pressure.sf.fb,hopkins:stellar.fb.winds,orr:ks.law,pandya:2021.loading.factors.of.fire}, that the average SFR and final stellar mass are approximately inversely proportional to the strength of feedback, which here is dominated by SNe with un-resolved cooling radii so essentially given by $p_{\rm term}$ and hence $\mathcal{F}_{{\bf v}}$. 

These results are entirely expected if we change $\mathcal{F}_{{\bf v}}$ systematically -- what is more interesting is if we look at how $\mathcal{F}_{{\bf v}}$ varies with $\langle v_{r} \rangle$. If we adopt the ``total momentum change'' formalism, we see a strong suppression of SF, particularly at late times. This is because, as discussed in \citet{hopkins:fire3.methods}, when stellar feedback normally would become less efficient in higher galaxy/halo masses and deeper potentials, the system will more and more often show gas in inflow around a SNe, which then strongly boosts its efficiency, so this decreasing feedback efficiency is partially offset. Changing, on the other hand, the residual thermal coupling of the gas has almost no effect, because at this resolution the cooling radii are unresolved.

Fig.~\ref{fig:mstar.mhalo} extends this to compare the stellar mass-halo mass relation from the same simulations, with halos masses spanning $\sim 10^{9}-10^{13}\,{\rm M_{\odot}}$ and stellar masses $\sim 10^{4}-10^{12}\,{\rm M_{\odot}}$. For halo masses $\lesssim 10^{11}\,{\rm M_{\odot}}$, we see weaker effects of $\mathcal{F}_{{\bf v}}$, for two reasons: (1) the resolution of these isolated dwarf zoom-ins is higher (see Table~1 in \citealt{hopkins:fire2.methods}), reaching $\sim 10^{2}\,{\rm M_{\odot}}$ for the smallest, so the cooling radii are increasingly resolved; and (2) because feedback in dwarfs is so strong, a much larger fraction of SNe explode in super-bubbles with $\langle v_{r} \rangle \gg 0$ (where $\mathcal{F}_{\bf v}$ is similar in all tested prescriptions). We also compare the results from FIRE-2 \citep{hopkins:fire2.methods}, which effectively assumed $\mathcal{F}_{\bf v} = 1$, but also a different numerical SNe implementation, UV background, cooling and thermo-chemistry, stellar evolution tables, and more; as well as the initial FIRE-3 runs from \citet[][H23]{hopkins:fire3.methods}, which adopted $\mathcal{F}_{\bf v}$ closer to the ``total momentum change'' model but are otherwise similar to our runs here (see \S~\ref{sec:compare}). In dwarfs, the difference between FIRE-2 and FIRE-3 is robust owing to the various changes above (and less sensitive to $\mathcal{F}_{\bf v}$, though it can be sensitive to the energy-conserving terms $\psi$; \citealt{hopkins:fire3.methods}), but we see that the substantial difference between FIRE-2 and the \citet{hopkins:fire3.methods} runs at halo masses $\gtrsim 10^{11}\,{\rm M_{\odot}}$ can primarily be attributed to the implicit choice of $\mathcal{F}_{\bf v}$. 

In the future it will be important to examine other parallel diagnostics which are sensitive to feedback properties. In Gandhi et al., in preparation, we will explore the stellar mass-metallicity relation (MZR), and its connection to the stellar mass-halo mass relation and satellite luminosity functions for dwarf galaxies, in more detail. Briefly, comparing the H23 simulations and those with $\mathcal{F}_{\bf v}=1$ in the MZR suggests that for the more massive galaxies, the first-order effect is to move along the MZR, but this is perhaps not surprising since they are in the very shallow, somewhat saturated (near-Solar) mass range. More notable differences appear comparing the FIRE-2 simulations to either FIRE-3 variant at the dwarf mass range (with low-mass dwarfs in FIRE-3 have higher metallicities, consistent with their less-efficient ``blowouts'' noted above), but we caution that those simulations used different yield tables and prompt Ia rates \citep[see discussion in][]{hopkins:fire3.methods}, so more work is needed to disentangle these effects. Similarly, the size-mass relation, galaxy stellar mass profiles, thick-versus-thin disk properties, metallicity gradients, and neutral gas velocity dispersions can be more sensitive probes of the details of stellar feedback couplings and should be studied to constrain these models \citep{orr:2020.resolved.dispersions.sfrs.correlations,parsotan:2021.mock.fire.profiles.z2,yu:2021.fire.bursty.sf.thick.disk.origin,rohr:2022.galaxy.halo.size.relation.fire,gurvich:2022.disk.settling.fire,klein:2024.size.mass.relation.lowmass.galaxies.with.mock.imaging.fire.agrees.perfectly.with.observations.but.must.model.obs,graf:2024.metal.gradients.mw.self.similar.upside.down.but.not.inside.out}.

\section{Conclusions}
\label{sec:conclude}

In galaxy formation simulations, it is common to treat SNe feedback with a subgrid model for kinetic and thermal injection representing the work done in any unresolved energy-conserving phase. We show that extending such treatments to be ``velocity-aware'' (i.e.\ dependent on the velocity field of gas surrounding the SNe, as required for strict energy conservation) introduces an ambiguity when SNe explode in converging flows. In this limit, there are a range of allowed (fully-conservative and analytically justifiable) sub-grid solutions, which differ strongly in how much momentum is imparted to the gas. This dependence is much stronger than the reasonably well-understood dependence of momentum coupled on ambient gas density, metallicity, or turbulence. We show that this can produce significant differences in cosmological galaxy-formation simulations, especially in high-mass galaxies at low resolution. We also stress that this is a {\em physical}, rather than numerical, ambiguity. As such, even in simulations which numerically couple feedback in a qualitatively different manner (e.g.\ ``delayed cooling'' models which do not explicitly couple momentum), the same fundamental ambiguity exists of how to scale mechanical feedback with surrounding gas velocities.

Given this intrinsic physical ambiguity and the absence of clear guidance from simulations which resolve individual SNe explosions in the relevant environments, we would recommend future work adopt the simplest possible model, namely our ``$\mathcal{F}_{{\bf v}}=1$'' formulation, as it introduces no additional implicit assumptions and is straightforward to compare to both previous simulations and idealized models. And as we showed this gives similar results in galaxy-formation simulations (at least for bulk galaxy properties) to the ``$\Delta$-Momentum'' formulation. But clearly, an important path for future work is to study the ambiguous cases here in more detail in higher-resolution simulations to develop more accurate sub-grid models for the extreme, but sometimes relevant limits. These would be analogous to classic experiments like those in \citet{kim.ostriker:sne.momentum.injection.sims}, but utilize inflow/outflow initial/boundary conditions (non-trivial but doable in Lagrangian codes like that used here, see \citealt{hopkins:gizmo.public.release}, but straightforward in Eulerian grid codes). We have shown that is important for such models to not simply quantify the final momentum state or ``total'' momentum change of the high-resolution simulation, as is usually done, but to think carefully about the {\em change} in this state specifically introduced at a given spatial, mass, and time scale by individual SNe as a function of their local properties, so that this can be implemented in sub-grid models without ``skipping over'' resolved MHD processes.

\begin{acknowledgements}
Support for PFH was provided by NSF Research Grants 1911233, 20009234, 2108318, NSF CAREER grant 1455342, NASA grants 80NSSC18K0562, HST-AR-15800. Numerical calculations were run on the Caltech compute cluster ``Wheeler,'' allocations AST21010 and AST20016 supported by the NSF and TACC, and NASA HEC SMD-16-7592.
\end{acknowledgements}

\bibliographystyle{mn2e}
\bibliography{ms_extracted}

\begin{appendix}

\section{Conservative Coupling of Mechanical Feedback}
\label{sec:methods}

\subsection{Coupling Equations}
\label{sec:methods:eqns}

Following \citet{hopkins:sne.methods} and \citet{hopkins:fire3.methods}, consider an ``event'' where a star particle ``$a$'' (with coordinate position ${\bf x}_{a}$, velocity ${\bf v}_{a}$, total particle/cell mass $m_{a}$) injects, in a given timestep $\Delta t$ (taking the code from initial time $t^{(0)}  \rightarrow t^{(1)} = t^{(0)} + \Delta t$), some ejecta mass $m_{{\rm ej},\,a}$ (which can be spread across different explicitly-followed species ``$s$,'' as $m_{{\rm ej},\,a} \equiv \sum_{s} m_{{\rm ej},\,a,\,s}$). In the rest-frame of the star particle, assume the ejecta is isotropic with total kinetic energy ${\rm KE}_{{\rm ej},\,a}$ (and some arbitrary set of scalar energies ``$j$'' including e.g.\ thermal and cosmic ray and radiation energies $U_{{\rm ej},\,a,\,j}$). The ejecta is  coupled to some set of gas cells ``$b$,'' which consist of all cells for which there is a non-vanishing oriented hydrodynamic face area ${\bf A}_{ba}$ that can be constructed between them and the star particle $a$ (this is defined rigorously in \citet{hopkins:sne.methods}, but essentially includes all cells $b$ which are either ``neighbors of $a$'' or for which ``$a$ is a neighbor of $b$''). Given a single event site $a$, then the update to the conserved quantities (e.g.\ $m$, $U$, and momentum ${\bf p}$) of cell $b$ is given by:
\begin{align}
\label{eqn:flux.m.coupling} m_{b,\,s}^{(1)} &= m_{b,\,s}^{(0)} + |\bar{\bf w}_{ba}|\,m_{{\rm ej},\,a,\,s}  \\ 
\nonumber U_{b,\,j}^{(1)} &= U_{b,\,j}^{(0)} + |\bar{\bf w}_{ba}|\,U_{{\rm ej},\,a,\,j}  \\ 
\nonumber {\bf p}_{b}^{(1)} &= {\bf p}_{b}^{(0)} + \Delta m_{ba}\,{\bf v}_{a} + \Delta {\bf p}_{ba}  \\ 
\nonumber \Delta {\bf p}_{ba} &\equiv \bar{\bf w}_{ba}\,p_{0,\,a}\ \ .
\end{align}
Here $\Delta m_{ba} \equiv |\bar{\bf w}_{ba}|\,m_{{\rm ej},\,a}$ and $\bar{\bf w}_{ba}$ is a normalized vector weight function defined in \citet{hopkins:sne.methods}.\footnote{Specifically, defining ${\bf x}_{ba} \equiv {\bf x}_{b}-{\bf x}_{a}$ and ${\bf A}_{ba}$ the oriented hydrodynamic area of intersection of faces between $b$ and $a$ \citep{hopkins:sne.methods}:
\begin{align}
\label{eqn:w.prime} w^{\prime}_{ba} &\equiv \frac{|\bar{\bf w}_{ba}|}{1+\Delta m_{ba}/m_{b}} \\
\nonumber \bar{\bf w}_{ba} &\equiv \frac{{\bf w}_{ba}}{\sum_{c}\,|{\bf w}_{ca}|} \\ 
\nonumber {\bf w}_{ba} &\equiv \omega_{ba}\, \sum_{+,\,-}\,\sum_{\alpha}\,(\hat{\bf x}_{ba}^{\pm})^{\alpha}\,\left( f_{\pm}^{\alpha} \right)_{a} \\ 
\nonumber \left( f_{\pm}^{\alpha} \right)_{a} &\equiv \left\{ \frac{1}{2}\,\left[1 +  \left( \frac{\sum_{c}\,\omega_{ca}\,|\hat{\bf x}_{ca}^{\mp}|^{\alpha}}{\sum_{c}\,\omega_{ca}\,|\hat{{\bf x}}_{ca}^{\pm}|^{\alpha}} \right)^{2}\right]\right\}^{1/2} \\
\nonumber (\hat{\bf x}^{+}_{ba})^{\alpha} &\equiv {|{\bf x}_{ba}|^{-1}}\,{\rm MAX}({\bf x}_{ba}^{\alpha},\,0)\,{\Bigr|}_{\alpha=x,\,y,\,z} \\
\nonumber (\hat{\bf x}^{-}_{ba})^{\alpha} &\equiv {|{\bf x}_{ba}|^{-1}}\,{\rm MIN}({\bf x}_{ba}^{\alpha},\,0)\,{\Bigr|}_{\alpha=x,\,y,\,z} \\
\nonumber \omega_{ba} & \equiv \frac{1}{2}\,\left(1-\frac{1}{\sqrt{1+({\bf A}_{ba}\cdot \hat{\bf x}_{ba})/(\pi\,|{\bf x}_{ba}|^{2})}}\right) \ \ .
\end{align}}
As shown therein, this ensures all the desired symmetries (e.g.\ isotropy in the frame of the star and $\sum_{b}  |\bar{\bf w}_{ba}| = 1$) as well as mass ($\sum_{b} \sum_{s} |\bar{\bf w}_{ba}|\,m_{{\rm ej},\,a,\,s} = \sum_{b} \Delta m_{ba} = m_{{\rm ej},\,a} $) and linear momentum conservation ($\sum_{b} \Delta {\bf p}_{ba} = \mathbf{0}$), with the total coupled radial momentum equal to $p_{0,\,a}$ in the star frame ($\sum_{b} |\Delta {\bf p}_{ba}| = p_{0,\,a}$).

\subsection{Enforcing Total Energy Conservation}
\label{sec:methods:egycon}

Without loss of generality, define
\begin{align}
\label{eqn:p0} p_{0,\,a} &\equiv \psi_{a}\,\chi_{a}\,( 2\,\epsilon_{a}\,m_{{\rm ej},\,a})^{1/2} 
\end{align}
in terms of some effective unit of kinetic energy
\begin{align}
\label{eqn:epsilon} \epsilon_{a} &\equiv f_{\rm kin}^{0}\,\mathcal{E}_{a} \equiv (1 - f_{U}^{0}) \mathcal{E}_{a}\, \\ 
\nonumber \mathcal{E}_{a} &\equiv  E_{{\rm ej},\,a} + \frac{1}{2}\sum_{b} m_{{\rm ej},\,a}\,w^{\prime}_{ba}\,|{\bf v}_{ba}|^{2}  
\end{align}
where $E_{{\rm ej},\,a}$ is the total ejecta energy (in the frame of the star, ignoring any surrounding gas), and $f_{U}^{0}$ and $f_{\rm kin}^{0} = 1-f_{U}^{0}$ are the fraction of the total energy in thermal (or other, e.g. radiation, cosmic rays, etc.) versus kinetic, fixed (hence the $^{0}$ superscript) to their values for e.g.\ an idealized \citet{sedov:book} solution in a homogeneous background, $f_{\rm kin}^{0} = 0.28$. At this stage, this is just a choice of units. As shown in \citet{hopkins:sne.methods}, the {\em exact}, discrete total energy conservation equation is:
\begin{align}
\label{eqn:egycon.reduced} \epsilon_{a} &= \sum_{b}\,\left[ \frac{|\Delta {\bf p}_{b}|^{2}}{2\,m_{b}\,(1+\mu_{b})}
+ \frac{{\bf v}_{ba} \cdot \Delta {\bf p}_{b}}{(1+\mu_{b})} 
\right] 
\end{align}
which becomes (inserting our definition of $\Delta {\bf p}_{b}$):
\begin{align}
\label{eqn:egycon} \frac{( \mathcal{E}_{a} - U_{a})}{\epsilon_{a}} &= (\psi_{a}\,\chi_{a})^{2}\,\beta_{2,\,a} + 2\,(\psi_{a}\,\chi_{a})\,\beta_{1,\,a} \\ 
\label{eqn:beta.1} \frac{\langle v_{r} \rangle_{a}}{v_{{\rm ej},\,a}} &\equiv \beta_{1,\,a} \equiv \left(\frac{m_{{\rm ej},\,a}}{2\,\epsilon_{a}}\right)^{1/2}\, \sum_{b} w^{\prime}_{ba}\, {{\bf v}_{ba} \cdot \hat{\bf w}_{ba}} \\
\label{eqn:beta.2}\frac{m_{{\rm ej},\,a}}{\langle M_{\rm coupled} \rangle_{a}} &\equiv \beta_{2,\,a} \equiv m_{{\rm ej},\,a}\,\sum_{b} \frac{w^{\prime}_{ba}\, |\bar{\bf w}_{ba}|}{m_{b}}
\end{align}
where ${\bf v}_{ba} \equiv {\bf v}_{b} - {\bf v}_{a}$.\footnote{We define $\langle v_{r} \rangle \equiv \beta_{1}\,v_{\rm ej} \equiv \sum_{b} w^{\prime}_{ba} {\bf v}_{ba} \cdot \hat{\bf w}_{ba}$ (with $v_{\rm ej} \equiv (m_{\rm ej}/2\,\epsilon_{a})^{1/2}$ some effective ejecta velocity) because it clearly reflects some weighted mean inflow/outflow velocity (projected along the direction of the coupled momentum) of the ambient gas in the star frame. Likewise $\langle M_{\rm coupled} \rangle \equiv m_{\rm ej}/\beta_{2} \equiv 1/\sum_{b} w^{\prime}_{ba}\,|\bar{\bf w}_{ba}|/m_{b}$ follows if we note that (for all resolutions of interest) $w^{\prime}_{ba} \approx |\bar{\bf w}_{ba}| \sim 1/N_{\rm ngb}$, where $N_{\rm ngb}$ is some effective number of neighbor cells to which the ejecta are coupled, so $\langle M_{\rm coupled} \rangle \sim 1/ \sum (N_{\rm ngb}^{-2}\,m_{b}^{-1}) \sim 1/(N_{\rm ngb} \times N_{\rm ngb}^{-2}\,\langle m_{b} \rangle ^{-1}) \sim N_{\rm ngb}\,\langle m_{b} \rangle$ represents some effectively-weighted gas mass to which the ejecta are coupled.}

Now, consider the case where the ejecta are in a purely energy-conserving Sedov-Taylor type phase, with some analytically-calculated (fixed/self-similar) ratio of kinetic to total energy $f_{\rm kin}^{0}$, where $\mathcal{E}_{a}$ represents the total energy available to the blastwave, and $U_{a}$ represents the fraction in non-kinetic sources, $U_{a} = f_{U}\,\mathcal{E}_{a} \rightarrow f_{U}^{0}\,\mathcal{E}_{a}$. Then clearly the left-hand side of Eq.~\ref{eqn:egycon} becomes unity ($\rightarrow 1$). Moreover given its strict degeneracy with $\psi_{a}$ we can take $\chi_{a} \rightarrow 1$ without loss of generality for now. It is then straightforward to solve and obtain the energy-conserving value of $\psi_{a}$:
\begin{align}
\psi_{a} &\equiv \frac{\sqrt{\beta_{2,\,a} + \beta_{1,\,a}^{2}} - \beta_{1\,a}}{\beta_{2,\,a}} \\
\nonumber &= \frac{1}{\beta_{2,\,a}^{1/2}}\left[ \sqrt{1 + \left(\frac{\beta_{1,\,a}}{\beta_{2,\,a}^{1/2} }\right)^{2}  } - \frac{\beta_{1,\,a}}{\beta_{2,\,a}^{1/2} } \right] \ .
\end{align}

The additional term $0 \le \chi_{a} \le 1$ then allows us to represent cases with less momentum coupled (more of the energy going into some other form). If we arbitrarily define some maximal or ``terminal'' momentum $p_{\rm term}$, then we simply impose:\footnote{Note that the ${\rm MIN}[1,\,\tilde{p}]$ function (with $\tilde{p} \equiv p_{\rm term} / \psi_{a}\,(2\,\epsilon_{a}\,m_{{\rm ej},\,a})^{1/2}$) in Eq.~\ref{eqn:chifun} is arbitrary and could be replaced with some other interpolation function ensuring $0\le \chi_{a} \le 1$, $\chi\rightarrow 1$ for $\tilde{p} \gg 1$, $\chi\rightarrow \tilde{p}$ for $\tilde{p} \ll 1$, if desired. But for our purposes, this has very small effects compared to the changes in $p_{\rm term}$ itself.} 
\begin{align}
\label{eqn:chifun} \chi_{a} &\rightarrow {\rm MIN}\left[ 1 \ , \  \frac{p_{\rm term}}{ \psi_{a}\,(2\,\epsilon_{a}\,m_{{\rm ej},\,a})^{1/2}  } \right]
\end{align}
which then determines the value of $f_{U} \ne f_{U}^{0}$, or $U_{a}$:
\begin{align}
\frac{U_{a}}{\mathcal{E}_{a}} &= 1 - \left[ (\psi_{a}\,\chi_{a})^{2}\,\beta_{2,\,a} + 2\,(\psi_{a}\,\chi_{a})\,\beta_{1,\,a} \right] \frac{\epsilon_{a}}{\mathcal{E}_{a}} \ .
\end{align}
This will ensure that the correct total energy is always coupled.\footnote{Our energy conservation equation is still a perfect equality even in the limit where we assume the ejecta and/shocked gas has cooled or entered some snowplow phase below our resolution scale. The only difference in e.g.\ the cooling limit is that energy has been transferred from a thermal reservoir to radiation. So we can keep the same derivation, thinking of $U$ as the thermal+radiated energy. It is not ``lost'' in this accounting -- we simply can decide if we wish to, at the end, explicitly add it to some radiation reservoir or ``deleted'' $U_{\rm rad}$. In fact, it is fine to couple it numerically as thermal energy in any case, given that in the only limit this could possibly apply, we are assuming $t_{\rm cool} \ll t_{\rm expand}$ of the ejecta shock, so it should be radiated away immediately (relative to the timescales resolved in-code), and if we have incorrectly under-estimated the cooling radius, the code can ``correct'' the hydrodynamics.}
We enforce the same timestepping and parallel communication/summation rules as described in \citet{hopkins:fire3.methods} (Appendix~B3) to ensure that multiple events acting on the same cell in the same timestep do not lead to spurious violations of conservation.

\subsection{Quantities Determining the Terminal Momentum}
\label{sec:methods:eval}

The terminal momentum $p_{\rm term}$ can be a function of various properties of the ambient gas and ejecta itself, e.g.\ 
\begin{align}
p_{\rm term} = p_{\rm term}(\langle \boldsymbol{\Upsilon}_{a} \rangle \ , \ \boldsymbol{\Psi}_{a} )
\end{align} 
where 
\begin{align}
\langle {\Upsilon}_{a,\,j} \rangle \equiv \sum_{b}  |\bar{\bf w}_{ba}|\,{\Upsilon}_{b,\,j}
\end{align} 
and $\boldsymbol{\Upsilon}_{b} = \{ n_{b},\, Z_{b},\, T_{b},\, ....\}$ is determined by various scalar properties of the neighboring gas cells, weighted appropriately by the fraction of the ejecta $|\bar{\bf w}_{ba}|$ associated with each cell (and $ \boldsymbol{\Psi}_{a}$ represents intrinsic ejecta properties such as $\mathcal{E}_{a}$, $m_{{\rm ej},\,a}$, etc.). Given that we are assuming isotropy in the frame of the star already above in our definition of weights to ensure e.g.\ linear momentum conservation, we use these weighted averages at the ejecta origin, rather than solving ``along cells'' or cones independently (which necessarily violates said isotropy). For $p_{\rm term}$ being a non-linear function of certain properties such as $n^{\alpha_{n}}\,Z^{\alpha_{Z}}$, we take $\sum_{b} |\bar{\bf w}_{ba}|\, n_{b}^{\alpha_{n}}\,Z_{b}^{\alpha_{Z}}$, to avoid pathological behaviors caused by covariances of these quantities.

\subsection{Practical Meaning of $\langle M_{\rm coupled} \rangle$}
\label{sec:methods:mcoupled}

As an aside, the parameter $\langle M_{\rm coupled} \rangle \equiv \left[ \sum_{b} w^{\prime}\,|\bar{\bf w}|_{ba}/m_{b} \right]^{-1} = \left[ \sum_{b} |\bar{\bf w}|^{2}_{ba}/(m_{b}  + |\bar{\bf w}|_{ba}\,m_{{\rm ej},\,a})\right]^{-1} \approx  \left[ \sum_{b} |\bar{\bf w}|^{2}_{ba}/m_{b}\right]^{-1}$ defined in Eq.~\ref{eqn:beta.2} provides a rigorous way of thinking of the total energy change of the system, but also about the effective total mass among which the ejecta energy and momentum are shared. This is especially useful for irregular-mesh or mesh-free methods such as those employed in codes like {\small GIZMO} or {\small AREPO} where the kernel into which SNe ejecta are deposited can include some neighbors with extremely small shared face areas ${\bf A}$. Notably given the quadratic dependence on the weight function $|\bar{\bf w}|^{2}_{ba}$, these will contribute negligibly to $\langle M_{\rm coupled} \rangle$. It is straightforward to estimate analytically, and verify directly in the simulations, given the definition of face \citep{hopkins:gizmo} and cubic spline kernel function for the volume decomposition used in the FIRE-3 {\small GIZMO} simulations plus the roughly-equal cell masses $m_{b} \approx \langle m_{b} \rangle$  presented here, that $\langle M_{\rm coupled} \rangle \sim (4-6) \times \langle m_{b} \rangle$, i.e.\ the effective coupled mass is a few times the median cell mass. This is quite different from some other implementations: for example the methods used in \citet{hu:photoelectric.heating,hu:2017.rad.fb.model.photoelectric,hu:2019.sne.models.like.fire.testing.calibration,lahen:2020.griffin.sims}, which adopt a different hydrodynamic method and face definition, kernel function, and method for depositing ejecta (attempting to directly resolve the energy-conserving phase rather than using a sub-grid model as discussed here), which give $\langle M_{\rm coupled} \rangle \sim (100-200) \times \langle m_{b} \rangle$. That, in turn, has an important consequence for the physical limit the coupling is in -- the \citet{hu:photoelectric.heating,hu:2017.rad.fb.model.photoelectric} method will therefore require $\sim 30\times$ better mass resolution (compared to the FIRE-2/3 method here) to begin to effectively capture the energy-conserving phase, in good agreement with the comparison of direct resolution tests presented in \citet{hopkins:sne.methods} and \citet{hu:photoelectric.heating,hu:2017.rad.fb.model.photoelectric}. Note there are potential advantages (e.g.\ smoother solutions, less sensitivity to inhomogeneous cell/particle distribution or small-scale phase structure) and disadvantaged (lower effective resolution, enhanced overcooling) to larger $\langle M_{\rm coupled} \rangle$; but regardless of the desired value it is useful to have a well-posed definition.

\section{Analytic Scalings for $\mathcal{F}_{{\bf v}}$}
\label{sec:analytic.fv}

\subsection{``Absolute'' Terminal Momentum Change}
\label{sec:analytic.fv:absolute}

Consider the rest-frame of a star which explodes isotropically in an isotropic, spherically-symmetric gaseous medium with a uniform radial velocity flow (i.e.\ for the region of interest around the star, the velocities are radial with some roughly-constant $v_{r}$), and let us make the shell approximation and assume the swept-up mass in the shell $M_{s} \gg m_{\rm ej}$ is much larger than the initial ejecta mass (which carried some energy $E_{0}$), and that the energy is divided between kinetic and thermal forms. All of these assumptions can be freed, we simply wish to illustrate a heuristic argument here. Energy conservation (pre-cooling) gives $E_{0} + (1/2)\,M_{s}\,v_{r}^{2} \approx (1/2)\,M_{s} v_{s}^{2} + U_{\rm th}$, where the thermal energy $U_{\rm th} \approx (3/2)\,(M_{s}/\mu)\,k_{B}\,T_{s} \approx f_{U}\,(1/2)\,M_{s}\,(v_{s} - v_{r})^{2}$ with some $f_{U}=1-f_{K}$ where $0\le f_{U} < 1$ is some similarity parameter whose exact value is not important. Re-arranging this, we can solve for $v_{s}(M_{s})=v_{r}\,f_{U}/(1+f_{U}) + \sqrt{2\,E_{0}/((1+f_{U})\,M_{s}) + v_{r}^{2}/(1+f_{U})^{2}}$ (or equivalently $M_{s}(v_{s})$). Now define the cooling time $t_{\rm cool} = [(3/2)\,k_{B}\,T_{s}] / [(\rho_{s}/\mu)\,\Lambda(T_{s})]$, where we will approximate $\Lambda \approx \Lambda_{0}\,(T_{s}/T_{0})^{\beta_{\Lambda}}$ over the dynamic range of interest and $\rho_{s} \approx \tilde{\rho}\,\rho$ for some shock compression ratio $\tilde{\rho}$. Define the expansion time $t_{\rm exp} = r/v_{s}$ and note $r=(M_{s}/ (4\pi/3)\,\rho)^{1/3}$, and it becomes possible to solve for $v_{s}$ (and hence $M_{s}$) at which $t_{\rm cool}(v_{s} = v_{s,\,c}) \approx t_{\rm exp}(v_{s} = v_{s,\,c})$. Then we define the terminal momentum as $p_{\rm term} \equiv M_{s} \,v_{s} - M_{s}\,v_{r}$ (i.e.\ the change in momentum of the swept-up material relative to its initial value). 

After some tedious algebra, the above derivation can be simplified to 
\begin{align}
p_{\rm term} \equiv \frac{E_{0}}{v_{\rm term}}\,\mathcal{F}_{{\bf v}}
\end{align} 
where $\mathcal{F}_{{\bf v}}$ satisfies: 
\begin{align}
\mathcal{F}_{{\bf v}} = [(1-y)^{7/3-2\beta_{\Lambda}} \{ 1- (1-f_{U})\,y/2\} ]^{3/(11-6\beta_{\Lambda})} 
\end{align}
where $y\equiv \mathcal{F}_{{\bf v}}\,\langle v_{r} \rangle/v_{\rm term}$. This must in general be solved numerically, but the limits are simple. For small $|\langle v_{r} \rangle|/v_{\rm term} \ll 1$, $\mathcal{F}_{{\bf v}} \rightarrow 1$ (as it should). For $\langle v_{r} \rangle \gg v_{\rm term}$, $\mathcal{F}_{{\bf v}} \rightarrow v_{\rm term}/\langle v_{r} \rangle$. And for $\langle v_{r} \rangle \ll -v_{\rm term}$, $\mathcal{F}_{{\bf v}} \rightarrow [(1-f_{U})/2]^{3}\,|\langle v_{r}\rangle/v_{\rm term}|^{10-6\,\beta_{\Lambda}}$. An approximate solution accurate to $\sim 10\%$ for $\langle v_{r} \rangle \ge 0$ is $\mathcal{F}_{{\bf v}} \approx 1/[1+(1-\psi)\,\langle v_{r} \rangle/v_{\rm term}]$ with $\psi \equiv 9\,(5+3\,f_{U})/[2\,(11-6\,\beta_{\Lambda})\,(9+\langle v_{r}\rangle^{2}/v_{\rm term}^{2})]$, and a (slightly less accurate but reasonable) approximation for $\langle v_{r} \rangle < 0$ is $\mathcal{F}_{{\bf v}} \approx 1+\tilde{v}_{-}+\tilde{v}_{-}^{2}+\tilde{v}_{-}^{3} + [(1-f_{U})/2]^{3}\,|\langle v_{r}\rangle/v_{\rm term}|^{10-6\,\beta_{\Lambda}}$ where $\tilde{v}_{-} \equiv (|\langle v_{r} \rangle|/v_{\rm term})\,([17-12\beta_{\Lambda}-3 f_{U}]/[22-12\beta_{\Lambda}])$. Here, $v_{\rm term} \propto E_{0}^{1/11}\,n^{2/11}\,\Lambda_{0}^{3/11}$ is\footnote{More precisely, assuming a strong-shock in fully-ionized primordial gas, we have $v_{\rm term} \sim 200\,{\rm km\,s^{-1}}\,(E_{0}/10^{51}\,{\rm erg})^{1/11}\,(n/{\rm cm^{-3}})^{2/11}\,(\Lambda_{0}/3\times10^{-23}\,{\rm erg\,s^{-1}\,cm^{3}})^{3/11}\,[(1+f_{U})^{10/11}\,f_{U}^{-3/11}/1.8]$, where $\Lambda_{0}$ is evaluated at $T=T_{0}=T_{s}(v_{s}=v_{s,\,c})$.} reasonably consistent with the simulation-calibrated $v_{\rm term}$ and its scalings $\mathcal{F}_{\mathcal{E}}$, $\mathcal{F}_{n}$, $\mathcal{F}_{Z}$ (where $Z$ appears via $\Lambda$). 

We can also simplify the above by taking $\beta_{\Lambda}\rightarrow -\infty$ mathematically, or equivalently say the cooling function $\Lambda(T)$ begins to rise steeply for decreasing $T$ below some critical temperature $T_{\rm cool}$, which in turn leads to rapid cooling when the post-shock temperature $T_{s} < T_{\rm cool}$, or equivalently, which the shock-ambient medium velocity $v_{s} - v_{\rm gas}$ falls below some critical ``cooling velocity'' $v_{\rm cool}$. We then simply calculate $p_{\rm term}$ when $v_{s}-v_{r} = v_{\rm cool}$, which gives $p_{\rm term} = (E_{0}/v_{\rm term})\,\mathcal{F}_{{\bf v}}$ with $\mathcal{F}_{{\bf v}} \rightarrow 1 / (1 + \langle v_{r} \rangle /v_{\rm term})$ (making the identification $v_{\rm term} = (1+f_{U})\,v_{\rm cool}/2$) for $v_{r} > -v_{\rm term}$ and $\mathcal{F}_{{\bf v}} \rightarrow \infty$ for $v_{r} \le -v_{\rm term}$. This has a simple closed form solution, and has the same limiting behaviors as above for $|\langle v_{r} \rangle| \ll v_{\rm term}$, $\langle v_{r} \rangle \gg v_{\rm term}$. While for $\langle v_{r} \rangle < -v_{\rm term}$ it formally diverges, the full expression above for finite $\beta_{\Lambda}$ effectively diverges as well (scaling for a more realistic $-1 \lesssim \beta_{\Lambda} \lesssim 1/2$ as $|\langle v_{r}\rangle/v_{\rm term}|^{\alpha_{v}}$ with $7 \lesssim \alpha_{v} \lesssim 16$ in this limit), and in practice the momentum coupled will always be restricted to $p_{\rm egy}$ in this limit.

\subsection{The ``$\Delta$-Momentum'' Imparted Per Event}
\label{sec:analytic.fv:delta}

Now suppose we wish to calculate the ``$\Delta$-Momentum'' as defined in the text -- i.e.\ the change in the final momentum of the system not relative to the initial state pre-SNe, but relative to the analytic final state {\em in the absence of a SNe}. In the analytic models here, we can define this as $\Delta p \equiv (M_{s}\,v_{s} - M_{s}\,v_{r})_{E_{0} = E_{0}} - (M_{s}\,v_{s} - M_{s}\,v_{r})_{E_{0} \rightarrow 0}$, so we need to subtract the solution in the limit of vanishing initial energy. We need to consider both cases (a) where the system does not efficiently cool so remains in the energy-conserving limit ($\chi=1$ and $p_{0} = M_{s}\,v_{s} - M_{s}\,v_{r} = \psi\,\sqrt{2\,m_{\rm ej}\,E_{0}}$) and (b) one where we are cooled and in the terminal limit $p_{0} = M_{s}\,v_{s} - M_{s}\,v_{r} = p_{\rm term} = (E_{0}/v_{\rm term})\,\mathcal{F}_{{\bf v}}$ above. 

For both energy-and-momentum-conserving limits (a) \&\ (b), if $v_{r} \ge 0$, it is easy to verify that $p_{0} \rightarrow 0$ as $E_{0} \rightarrow 0$ given our expressions for $\psi\,\sqrt{2\,m_{\rm ej}\,\epsilon}$ and $p_{\rm term} = (E_{0}/v_{\rm term})\,\mathcal{F}_{{\bf v}}$ above (here both the dimensional momentum goes to zero, but also the dimensionless pre-factors $\psi$ and $\mathcal{F}_{{\bf v}}$ go to zero if $\langle v_{r} \rangle > 0$, because $\langle v_{r} \rangle / (v_{\rm ej},\,v_{\rm term}) \rightarrow +\infty$). So the ``$\Delta$-Momentum'' is identical to the $p_{\rm term}$ we already derived
\begin{align}
\Delta p_{\rm egy,\,term} = p_{0,\,{\rm egy,\,term}}\ \ \ \ \ (\langle v_{r} \rangle \ge 0)\ .
\end{align}

For $\langle v_{r} \rangle < 0$, noting that $\langle v_{r} \rangle / v_{\rm ej} = \langle v_{r} \rangle \,\sqrt{m_{\rm ej}/2\,\epsilon} \rightarrow -\infty $ as $E_{0} \rightarrow 0$, case (a) gives $\psi\rightarrow -2\,\beta_{1}/\beta_{2}$ and $p_{0} \rightarrow -2\,M_{\rm coupled}\,\langle v_{r} \rangle$ in this limit, independent of $E_{0}$ (\S~\ref{sec:methods:limits}). Subtracting this, we obtain $\Delta p = p_{0}(E_{0})-p_{0}(E_{0}\rightarrow 0) = M_{\rm coupled}\,v_{r} + \sqrt{(M_{\rm coupled}\,v_{r})^{2} + 2\,M_{\rm coupled}\,\epsilon}$, equivalent to taking $\psi\rightarrow \psi^{\prime} = \beta_{2}^{-1/2}[ \sqrt{1 +(|\beta_{1}/\beta_{2}^{1/2}|)^{2} } - |\beta_{1}/\beta_{2}^{1/2}| ]$ (identical to the expression for $\beta_{1} > 0$, with the absolute values now in place), 
\begin{align}
\Delta p_{\rm egy} &= \psi(\beta_{1}\rightarrow |\beta_{1}|)\,p_{\rm ej} \ \ \ \ \ (\langle v_{r} \rangle < 0) \\
\nonumber &= \sqrt{2 \langle M_{\rm coupled} \rangle \epsilon}\,\left[\sqrt{1 + \frac{|\beta_{1}|^{2}}{\beta_{2}}}  - \frac{|\beta_{1}|}{\beta_{2}^{1/2}}\right] \ . 
\end{align}
Case (b) must again be solved numerically in general, but noting that for $E_{0} \rightarrow 0$, $\langle v_{r} \rangle/v_{\rm term} \rightarrow -\infty$, the previous limiting solution for $\mathcal{F}_{{\bf v}}\rightarrow \mathcal{F}_{{\bf v},\,0} = [(1-f_{U})/2]^{3}\,|\langle v_{r} \rangle/v_{\rm term}|^{10-6\beta_{\Lambda}}$ applies and we can write $\mathcal{F}_{{\bf v}} \rightarrow \mathcal{F}_{{\bf v},\,0} + \mathcal{F}_{{\bf v},\,\Delta}$ and obtain $\mathcal{F}_{{\bf v},\,\Delta}$. If desired, this can be well-approximated for $\langle v_{r} \rangle < 0$ and $\tilde{v}\equiv -\langle v_{r} \rangle/v_{\rm term}  \ll 1$ by $\mathcal{F}_{{\bf v},\,\Delta} \rightarrow 1/(1-\tilde{v}_{1})$ with $\tilde{v}_{1} \equiv \tilde{v}\,(17-12\beta_{\Lambda}-3 f_{U})/(22-12\beta_{\Lambda})$ and for $\tilde{v} \gg 1$ by $\mathcal{F}_{{\bf v},\,\Delta} \rightarrow 1/\tilde{v}_{2}$ with $\tilde{v}_{2} \equiv \tilde{v} / [7-6\beta_{\Lambda} + 6/(1-f_{U})]$. For intermediate values and $\beta_{\Lambda} \gtrsim -1$ the interpolation function $\mathcal{F}_{{\bf v},\,\Delta} \sim (1 + \tilde{v}_{1} + \tilde{v}_{1}^{2} + \tilde{v}_{1}^{3}) / [1 +(\tilde{v}_{1}^{2} + \tilde{v}_{1}^{3})\,\tilde{v}_{2}]$ is acceptable,
\begin{align}
\nonumber \Delta p_{\rm term} &= \frac{E_{0}}{v_{\rm term}}\,\mathcal{F}_{{\bf v},\,\Delta} \\ 
\mathcal{F}_{{\bf v},\,\Delta} &\sim \frac{1 + \tilde{v}_{1} + \tilde{v}_{1}^{2} + \tilde{v}_{1}^{3}}{1 +(\tilde{v}_{1}^{2} + \tilde{v}_{1}^{3})\,\tilde{v}_{2}}\ \ \ \ \ (\langle v_{r} \rangle < 0) \ . 
\end{align}

Note that the behavior in case (a) for $\langle v_{r} \rangle$, though derived for an energy-conserving limit, can be trivially written into our standard formulation from \S~\ref{sec:methods} by defining $\mathcal{F}_{{\bf v}} \rightarrow {\rm MIN}[\mathcal{F}_{{\bf v},\,\Delta} \ , \ \Delta p_{\rm egy}\,v_{\rm term}/E_{0} ]$ for $\langle v_{r} \rangle < 0$.

\section{Previous FIRE Treatments}
\label{sec:compare} 

It is worth clarifying the difference between FIRE-1 \citep{hopkins:2013.fire}, FIRE-2 \citep{hopkins:fire2.methods}, and the initial version of FIRE-3 (``3.0'') in \citet{hopkins:fire3.methods}, in the SNe coupling specifically. All adopted a qualitatively similar hybrid energy-momentum coupling as Eq.~\ref{eqn:p.coupled.summary}, $p_{0} = {\rm MIN}[p_{\rm egy},\,p_{\rm term}]$. The scalings for $p_{t,\,0}$, $\mathcal{F}_{\mathcal{E}}$, $\mathcal{F}_{n}$, and $\mathcal{F}_{Z}$ in $p_{\rm term}$ are almost identical between 1/2/3.0/here, and weak. FIRE-1 adopted a much simpler, scalar kernel mass-weight for deposition, while FIRE-2 \&\ 3.0 adopted a normalized vector weight scheme as we do (\S~\ref{sec:methods:eqns}); the consequences of these differences are the subject of \citet{hopkins:fire2.methods}. FIRE-1 \&\ 2 effectively took $\mathcal{F}_{{\bf v}}$ \&\ $\psi$ to be constant, while 3.0 and here vary $\mathcal{F}_{{\bf v}}$ and $\psi$ to ensure energy conservation. There are two significant differences between FIRE-3.0 and here.\footnote{In \citet{hopkins:fire3.methods}, Appendix~B (Eqs.~B13-B19 therein), a rather complicated set of equations is presented for $\psi_{ba}\,\chi_{ba}$ (and slightly different notation was used), but we can simplify these for the limits of interest here (e.g.\ $m_{b} \gg m_{\rm ej}$). Doing so, for $m_{b}<m_{t,\,b} \sim p_{{\rm term},\,0,\,ba}^{2}/2\,\epsilon_{a}$, $\chi_{ba}\rightarrow 1$ with $\psi_{ba} \sim \sqrt{m_{b}/\Delta m_{ba}}/(1+2\,{\rm MAX}[0\, , \,v_{\rm ej}^{-1} \sum_{b} w^{\prime \,1/2}_{ba}\,{\bf v}_{ba} \cdot \hat{\bf w}_{ba}])$, and for $m_{b} \ge m_{t,\,b}$,  $\chi_{ba} = {\rm MIN}[p_{{\rm term},\,ba} / p_{{\rm egy},\,ba} \ , \ 1]$ as usual, with $p_{\rm term} \rightarrow |\bar{\bf w}_{ba}|^{-1/2}\,p_{{\rm term},\,0}\,{\rm MIN}[ 1 \ , \ v_{\rm term}/\langle v_{r} \rangle]$. So for low resolution and $\langle v_{r} \rangle \lesssim 0$, $p_{0,\,ba} \rightarrow |\bar{\bf w}_{ba}|^{-1/2}\,{\rm MIN}[p_{{\rm term},\,ba} \ , \ \sqrt{2\,\epsilon\,m_{b}} ] \sim |\bar{\bf w}_{ba}|^{-1/2}\,p_{{\rm term},\,ba}$, i.e.\ $F_{v} \rightarrow |\bar{\bf w}_{ba}|^{-1/2}$.}

{\bf (1)} In FIRE-3.0, we implicitly assumed something like the ``terminal velocity'' $v_{\rm term}$/$v_{\rm cool}$ model in \S~\ref{sec:analytic.fv:absolute} for $\mathcal{F}_{{\bf v}}$, giving $\mathcal{F}_{{\bf v}}$ rising with decreasing $\langle v_{r}\rangle$ to an imposed maximum of $\sim |\bar{\bf w}_{ba}|^{-1/2}$. This arose because $p_{\rm term}$ was written in terms of some $v_{\rm cool}$, which is equivalent to other ways of writing it when $\langle v_{r} \rangle = 0$ but, as we showed, differs for $\langle v_{r} \rangle \ll 0$. As noted in the text, this is the main difference between the massive-galaxy SF histories therein and those in FIRE-2 (which took $\mathcal{F}_{{\bf v}} = 1$). 

{\bf (2)} In FIRE-1/2/3.0, we applied solutions ``in cones.'' For example in 3.0, we took $\Delta {\bf p}_{ba} \equiv \bar{\bf w}_{ba}\,\psi_{a}\,\chi_{a}\,(2\,\epsilon_{a}\,m_{\rm ej,\,a})^{1/2} \rightarrow \bar{\bf w}_{ba}\, \psi^{\prime}_{ba}\, \chi^{\prime}_{ba}\,(2\,\epsilon_{a}\,m_{\rm ej,\,a})^{1/2}$ with the function $\psi^{\prime}_{ba}\,\chi^{\prime}_{ba}$ evaluated within each neighbor $b$ instead of being uniform for the entire coupled group. This was done with a desire to represent the highest-possible resolution solution and avoid imposing spherical symmetry on the surrounding gas around each SNe, but it introduces a couple of subtle issues. First, it violates the machine-accurate isotropy and linear momentum conservation around each SNe, which was discussed in \citet{hopkins:sne.methods} but there we found the effect was small and the method still avoided the more severe pathological effects of previous FIRE-1 like implementations. Second, this has the unintended effect of ``up-weighting'' the tails or outer-edges of the coupling kernel, because the effective momentum-per-cell in e.g.\ the energy-conserving limit scales as $\sim |\bar{\bf w}_{ba}|\,\sqrt{m_{b}/\Delta m_{ba}} \propto |\bar{\bf w}_{ba}|^{1/2}$. This in turn makes the results more sensitive to the coupling kernel, and makes the effective ``coupled mass'' $\langle M_{\rm coupled} \rangle$ much larger,\footnote{Formally, $\langle M_{\rm coupled} \rangle \sim [\sum_{b} |\bar{\bf w}_{ba}|^{1/2}\,m_{b}^{1/2}]^{2}$, instead of $\langle M_{\rm coupled} \rangle \sim [\sum_{b} |\bar{\bf w}_{ba}|^{2}\,m_{b}^{-1}]^{-1}$ as herein. In our  tests the former is factor $\sim 3$ larger.} which erases any effective resolution gain.

\end{appendix}

 \, \\

\end{document}